\documentclass[aps,showpacs,pra,twocolumn,groupedaddress]{revtex4-1}
\usepackage{graphicx}

\begin{document}

\title{Non-Markovian dynamics of a damped driven two-state system}

\author{P. Haikka}
\email[]{pmehai@utu.fi}
\homepage[]{www.openq.fi}
\affiliation{Turku Centre for Quantum Physics, Department of Physics and Astronomy, University of Turku, 20014 Turku, Finland}

\author{S. Maniscalco}
\email[]{sabrina.maniscalco@utu.fi}
\homepage[]{www.openq.fi}

\affiliation{Turku Centre for Quantum Physics, Department of Physics and Astronomy, University of Turku, 20014 Turku, Finland}
\date{\today}

\begin{abstract}
We study a driven two-state system interacting with a generic structured environment. We outline the derivation of the time-local microscopic non-Markovian master equation, in the limit of weak coupling between the system and the reservoir, and we derive its analytic solution for general reservoir spectra in the regime of validity of the secular approximation. We also consider the non-Markovian master equation without the secular approximation and study the effect of nonsecular terms on the system dynamics for two classes of reservoir spectra: the Ohmic and the Lorentzian reservoir. Finally, we derive the analytic conditions for complete positivity of the dynamical map, with and without the secular approximation. Interestingly, the complete positivity conditions have a transparent physical interpretation in terms of the characteristic time-scales of phase diffusion and relaxation processes. 
\end{abstract}

\maketitle

\section{Introduction}
All quantum systems are open, i.e., they interact with an environment. The interaction with the environment leads to dissipation and decoherence due to a flow of energy and/or information from the system to the environment \cite{breuer&petruccione, weiss}. The dynamics of dissipation and decoherence in an open quantum system depends on the properties of the environment, and therefore can be altered by modifying characteristics of the environment such as its spectrum \cite{Bollinger09,Maniscalco04a}.

A common example of environment is the quantized electromagnetic field, typically modeled as an infinite chain of non-interacting quantum harmonic oscillators. The coupling of the quantum system to the environment is described by the spectral density function. If the system couples to all modes of the environment in an equal way the spectrum of the reservoir is flat. If, instead, the spectral density function strongly varies with the frequency of the environmental oscillators, the environment is said to be structured. Structured environments arise in many physical situations, e.g., in photonic band-gap materials and lossy optical cavities \cite{lambropoulos, haroche}. In these systems the reservoir memory effects induce a feedback of information from the environment into the system. We call these systems non-Markovian \cite{JPRL09}.

In this paper we study, to second order in perturbation theory with respect to system-reservoir coupling, the non-Markovian dynamics of a driven two-state system in presence of a structured reservoir with a generic spectral density.  One of the first studies on the general dynamical properties of this model dates back to almost twenty years ago, when Lewenstein and Mossberg studied a driven atom inside both an optical cavity with a Lorentzian spectral density and a microwave cavity with a step function spectral density \cite{lewenstein}. 
Further studies on the laser-induced modification of the spontaneous decay of an atom embedded in a structured reservoir are given in Refs. \cite{brinati,janowicz,fanchini,budini,tanas1,tanas2}.

We extend these results in several ways. First of all we present the time-local non-Markovian master equation for the dynamics and its solution, valid in the limit of weak couling between the system and the environment. The memory effects due to the reservoir structure are contained in time-dependent decay rates. We show that, for short initial times, the dependence of the decay rates on the driving laser is more complicated than the one presented in Ref.  \cite{lewenstein}. For these short initial times, indeed, the spontaneous decay of the atom can not only be suppressed or enhanced, but also partly reversed, when non-Markovian oscillations induced by reservoir memory effects are present.

The importance of the driven two-state model is especially pronounced in quantum computation and quantum technologies, where one or more driven qubits constitute the basic building block of quantum logic gates \cite{nilsen}. Different implementations of qubits for quantum logic gates are subjected to different types of environmental noise, i.e., to different environmental spectra \cite{Bollinger09}. In this study we focus on two examples of a structured reservoir, namely the Ohmic and the Lorentzian reservoirs. Owing to the microscopic approach that we adopt in this paper, we can make a comparison of the microscopic processes underlying the system dynamics for the two different reservoirs. This knowledge can aid the search for physical set-ups that best retain quantum properties under dissipative dynamics.

We study the system dynamics with and without the widely used secular approximation, singling out its limits of validity. The time-scale for nonsecular phenomena ranges typically from small to intermediate in comparison with the time-scale for relaxation processes and therefore the secular approximation may not be consistent with studies of non-Markovian dynamics. A recent article by Cummings and Hu further elucidates the importance of nonsecular studies on open quantum systems \cite{hu}.

Our investigation of the effects of nonsecular terms on the dynamics of the driven two-state system brings to light the existence of nonsecular oscillations in the population of the two-state system. These oscillations have similar nature to those observed in the entanglement dynamics in Ref. \cite{vasile}. Contrarily to the oscillations due to non-Markovianity, nonsecular oscillations persist for times much longer than the reservoir correlation time. Moreover, our analysis shows that the nonsecular terms affect also the asymptotic long time values of the Bloch vector components, i.e., their impact exceeds the time-scale of nonsecular oscillations.

Finally, an important result we present in the paper is the analysis of conditions for complete positivity (CP) of the system. All phenomenological or approximated non-Markovian master equations may lead to unphysical results for certain values of the parameters. In order to guarantee the physicality of the solution of the master equation, one needs to study the complete positivity of the dynamical map. This is by no means an easy task. Explicit conditions for CP have been up to now obtained only for very simple systems \cite{Maniscalco07,Breuer09}. Here we study for the first time the CP conditions for the non-Markovian driven two-state system and show that they have a clear physical interpretation in terms of the decay rates.

The paper is organized as follows. In Sec. \ref{the model} we introduce the microscopic Hamiltonian model and the non-Markovian nonsecular time-local master equation describing the dynamics of the driven two-level atom in a generic structured reservoir. In Sec. \ref{decay rates} we derive the analytic expressions of the non-Markovian time-dependent decay rates for the special cases of a Lorentzian and an Ohmic reservoir and we discuss the physical processes characterizing the system dynamics. These results are used to study the solutions of the optical Bloch equations in Sec. \ref{dynamics} in the secular and the non-secular regime. In Sec. \ref{cp} we derive the necessary and sufficient conditions for completely positivity. Finally Sec. \ref{conclusions} summarizes the results and presents conclusions.

\date{\today}

\section{The microscopic model}\label{the model}

We consider a two-level atom with Bohr frequency $\omega_A$ interacting with a driving laser of frequency $\omega_L$ almost resonant with the atomic transition, i.e., $|\Delta|=|\omega_A-\omega_L|\ll\omega_A$. The two-level atom is embedded in a zero-T thermal bosonic reservoir modeled by an infinite chain of quantum harmonic oscillators. In a frame rotating with the laser frequency $\omega_L$ the total Hamiltonian for this system, in units of $\hbar$, is given by
\begin{equation}
\label{eq:h}
H=H_S+H_E+H_I,
\end{equation}
where
\begin{eqnarray}
\label{eq:hs}
H_S&=&\frac{1}{2}(\Delta\sigma_z+\Omega\sigma_x),\\ \label{eq:he}
H_E&=&\sum_k\omega_ka_k^{\dagger}a_k,\\ \label{eq:hi}
H_I&=&\sum_k g_ke^{-i\omega_L t}a_k^{\dagger}\sigma_-+g_k^*e^{i \omega_L t}a_k\sigma_+,
\end{eqnarray}
are the free Hamiltonians of the system and the environment and the interaction Hamiltonian, respectively, $\sigma_{x,y,z}$ are the Pauli operators, $\sigma_{\pm}$ the atomic inversion operators and $a_k$ the annihilation operator of quanta in the reservoir $k$-th mode.\\
The Rabi frequency $\Omega$ describes the strength of the interaction between the atom and the laser and it is taken to be small compared to the atomic and laser frequencies, $\Omega\ll\omega_A,\omega_L$. The interaction strength between the two-level atom and the $k$-th mode of the reservoir is given by $g_k$. In the limit of a continuum of reservoir modes $\sum_k|g_k|^2\rightarrow\int d\omega J(\omega),$ where $J(\omega)$ is the spectral density function, characterizing the reservoir spectrum. In this paper we focus on structured reservoirs, i.e., reservoirs with a spectrum that varies sensibly with the environmental oscillators frequency.\\
The description of a quantum system in a structured reservoir requires  non-Markovian approaches since the reservoir correlation time is typically longer than other time-scales of the system dynamics. In the following subsection we present the microscopic non-Markovian master equation for the system introduced above. We will see how useful information on the system dynamics can be inferred already by looking at the form of the master equation and in particular by studying the behavior of the time-dependent decay rates appearing in the equations. 

\subsection{Time-local master equation}
We use the time-convolutionless (TCL) projection operator technique to obtain the master equation for the driven two-level atom starting from the microscopic model of Eqs. (\ref{eq:h})-(\ref{eq:hi}) \cite{breuer&petruccione}. In the limit of weak coupling between the system and the environment the TCL generator is expanded to second order with respect to a coupling constant quantifying the strength of the interaction between the system and the environment. In Ref. \cite{haikka} one of us has demonstrated that the time-local non-Markovian master equation describing the system under study can be written in the form
\begin{equation}
\label{master equation}
\frac{d\bar{\rho}(t)}{dt}=-i[\bar{H}_S+\bar{H}_{LS},\bar{\rho}(t)] +\mathcal{D}[\bar{\rho}(t)]+\mathcal{D}'[\bar{\rho}(t)],
\end{equation}
where bars indicate that the operators are given in the dressed state basis $|\psi_\pm\rangle=\pm\sqrt{C_{\pm}}|e\rangle+\sqrt{C_{\mp}}|g\rangle$, where $|e\rangle$ and $|g\rangle$ are the atomic excited and ground state, and the coefficients $C_\pm$ are 
\begin{equation}
\label{cpm}
C_{\pm}=\frac{\Delta\pm\omega}{2\omega},\quad C_0=\frac{\Omega}{2\omega},
\end{equation}
with
\begin{equation}
\label{omega}
\omega=\sqrt{\Delta^2+\Omega^2}
\end{equation}
the energy separation between the eigenstates of the driven atom.
The unitary part of Eq. (\ref{master equation}) is governed by the Hamiltonians 
\begin{eqnarray}
\label{dissipator}
\bar{H}_S &=& \frac{\omega}{2}\bar{\sigma}_z, \\
\bar{H}_{LS} &=& \lambda_-(t)C_-^2\bar{\sigma}_-\bar{\sigma}_++\lambda_+(t)C_+^2\bar{\sigma}_+\bar{\sigma}_- \nonumber \\
&+&\lambda_0(t)C_0^2\bar{\sigma}_z^2, \label{eq:HLS}
\end{eqnarray}
namely the free Hamiltonian and the Lamb shift Hamiltonian, respectively. The latter one describes a small shift in the energy of the eigenstates of the two-level atom. This term has no qualitative effect on the dynamics of the system and therefore will be neglected in the following. \\
The dissipator in Eq.  (\ref{master equation})  has been written as the sum of two terms, $\mathcal{D}$ and $\mathcal{D}'$. The first term is
\begin{eqnarray}
\label{secular dissipator}
\mathcal{D}[\bar{\rho}(t)]&=&C_+^2\gamma_+(t)\left[\bar{\sigma}_-\bar{\rho}(t)\bar{\sigma}_+-\frac{1}{2}\{\bar{\sigma}_+\bar{\sigma}_-,\bar{\rho}(t)\}\right]\nonumber\\
&+&C_-^2\gamma_-(t)\left[\bar{\sigma}_+\bar{\rho}(t)\bar{\sigma}_--\frac{1}{2}\{\bar{\sigma}_-\bar{\sigma}_+,\bar{\rho}(t)\}\right]\nonumber\\
&+&C_0^2\gamma_0(t)\left[\bar{\sigma}_z\bar{\rho}(t)\bar{\sigma}_z-\frac{1}{2}\{\bar{\sigma}_z\bar{\sigma}_z,\bar{\rho}(t)\}\right].
\end{eqnarray}
The second term has a more complicated form and
contains the contribution of the so-called nonsecular terms, i.e., terms oscillating rapidly with respect to the dressed atom characteristic time $\tau_S=\omega^{-1}$,
\begin{eqnarray}
\mathcal{D}'[\bar{\rho}(t)]\!\!&=&\!\!\!\left[\frac{\gamma_-(t)}{2}-i\lambda_-(t)\right]\!\!\Big\{C_-C_0\big[\bar{\sigma}_+\bar{\rho}(t)\bar{\sigma}_z-\bar{\sigma}_z\bar{\sigma}_+\bar{\rho}(t)\big] \nonumber \\
&+& C_+C_-\big[\bar{\sigma}_+\bar{\rho}(t)\bar{\sigma}_+-\bar{\sigma}_+\bar{\sigma}_+\bar{\rho}(t)\big]\Big\}\nonumber\\
&+&\!\!\!\left[\frac{\gamma_+(t)}{2}-i\lambda_+(t)\right]\!\!\Big\{C_+C_0\big[\bar{\sigma}_-\bar{\rho}(t)\bar{\sigma}_z-\bar{\sigma}_z\bar{\sigma}_-\bar{\rho}(t)\big] \nonumber \\
&+&C_+C_-\big[\bar{\sigma}_-\bar{\rho}(t)\bar{\sigma}_--\bar{\sigma}_-\bar{\sigma}_-\bar{\rho}(t)\big]\Big\}\nonumber\\
&+&\!\!\!\left[\frac{\gamma_0(t)}{2}-i\lambda_0(t)\right]\!\!\Big\{C_-C_0\big[\bar{\sigma}_z\bar{\rho}(t)\bar{\sigma}_--\bar{\sigma}_-\bar{\sigma}_z\bar{\rho}(t)\big] \nonumber \\
&+&C_+C_0\big[\bar{\sigma}_z\bar{\rho}(t)\bar{\sigma}_+-\bar{\sigma}_+\bar{\sigma}_z\bar{\rho}(t)\big]\Big\}+ h.c.\label{dissipator'}
\end{eqnarray}
where $h.c.$ denotes Hermitian conjugation.\\ 
As for all time-local master equations, the non-Markovian effects are contained in the coefficients $\gamma_{\xi}(t)$ and $\lambda_\xi(t)$, with $\xi\in\{-,0,+\}$, which arise from the real and imaginary part of the reservoir correlation function, respectively \cite{haikka}. These coefficients read
\begin{eqnarray}
\label{rates}
\gamma_{\xi}(t)&=&2\int_0^td\tau\int d\tilde{\omega}J(\tilde{\omega})\cos[(\omega_L+\xi\omega-\tilde{\omega})\tau],\\
\lambda_{\xi}(t)&=&\int_0^td\tau\int d\tilde{\omega}J(\tilde{\omega})\sin[(\omega_L+\xi\omega-\tilde{\omega})\tau]. \label{lambs}
\end{eqnarray}
For times longer than the reservoir correlation time $\tau_C$ the decay rates attain their stationary Markovian values $\gamma_{\xi}^M\equiv\lim_{t\rightarrow\infty}\gamma_{\xi}(t)$ and $\lambda_{\xi}^M\equiv\lim_{t\rightarrow\infty}\lambda_{\xi}(t)$. Therefore the first non-Markovian corrections on the dynamics of the driven two-level atom are visible only for small initial times $t \simeq \mathcal{O}(\tau_C)$.\\ 
We observe that the dynamics of the driven two-level atom comprises of three different dynamical effects occurring at three different respective time-scales. Indeed, the dynamics of both dissipation and decoherence occur at a time-scale of the order of the relaxation time-scale $\tau_R$, which is defined by the properties of the reservoir. Nonsecular terms cause oscillations, occurring over the typical time-scale of the system, $\tau_S=\omega^{-1}=(\Delta^2+\Omega^2)^{-1/2}$, for the driven two-level atom. Finally, as mentioned above, the non-Markovian memory effects happen for times shorter or of the order of the reservoir correlation time-scale $\tau_C$.\\

\subsection{The secular approximation}
Conventionally the nonsecular terms, contained in the dissipator $\mathcal{D}'$, are neglected in the secular approximation when $\tau_S \ll \tau_R$ \cite{breuer&petruccione}.  However, as one might expect, a non-Markovian description of the short-time dynamics is often incompatible with the secular approximation. In order to investigate the effects of the nonsecular terms on the non-Markovian dynamics, we focus instead on two regimes identified by the mutual relationship between the characteristic time-scale $\tau_S$ and the reservoir correlation time-scale $\tau_C$.\\
The first regime we call the {\sl secular regime}, characterized by the condition $\tau_S\ll\tau_C$. In this regime the nonsecular terms are negligible even at short non-Markovian time scales and we can make the secular approximation. The second regime is the {\sl nonsecular regime}, characterized by the opposite condition, i.e., $\tau_C\ll\tau_S$. In this case we must retain the nonsecular terms to correctly describe the non-Markovian dynamics.\\
When the secular approximation is valid, and the dissipator $\mathcal{D}'$ can be neglected, the coefficients $\lambda_\xi(t)$ appear only in the Lamb-shift Hamiltonian of Eq. (\ref{eq:HLS}), and therefore they describe a time-dependent renormalization of the dressed atomic energy. Moreover, the master equation is in time-dependent Lindblad form and the coefficients $\gamma_{\xi}(t)$ are proportional to the decay rates associated to transitions between the atomic dressed states described by the operators $\bar{\sigma}_+$,  $\bar{\sigma}_-$ and  $\bar{\sigma}_z$.
In this case, the dynamics of the system can be inferred from the time evolution of the decay rates for different types of reservoirs in terms of direct and reversed quantum jumps between dressed states, as suggested by the Non-Markovian quantum jump (NMQJ) method of Ref. \cite{nmqj,nmqj2}. We will discuss this point further in Sec. III.\\
When the secular approximation is not valid, the master equation is not in Lindblad-type form. Both $\gamma_{\xi}(t)$ and $\lambda_\xi(t)$ appear now in the dissipator $\mathcal{D}'$. In this case it is not possible to extract from the master equation the jump operators, describing transitions between the dressed states, and the associated time-dependent decay rates. However, as we will see in Sec. IV, the nonsecular terms give rise to interesting effects not only at intermediate times but also in the asymptotic long time regime. 

\section{Time-dependent coefficients and NMQJ interpretation} \label{decay rates}

In the following we specify our study to two exemplary reservoir spectra widely used in the literature, namely the Lorentzian and the Ohmic spectra. Our aim is to investigate how both the system dynamics and the validity of the secular approximation depend on the properties of the reservoir spectrum. We give analytic expression for the time-dependent coefficients and use the NMQJ method to compare the microscopic dynamics of the driven two-state system in the two exemplary reservoirs.

\begin{figure*}
\includegraphics[width=15cm]{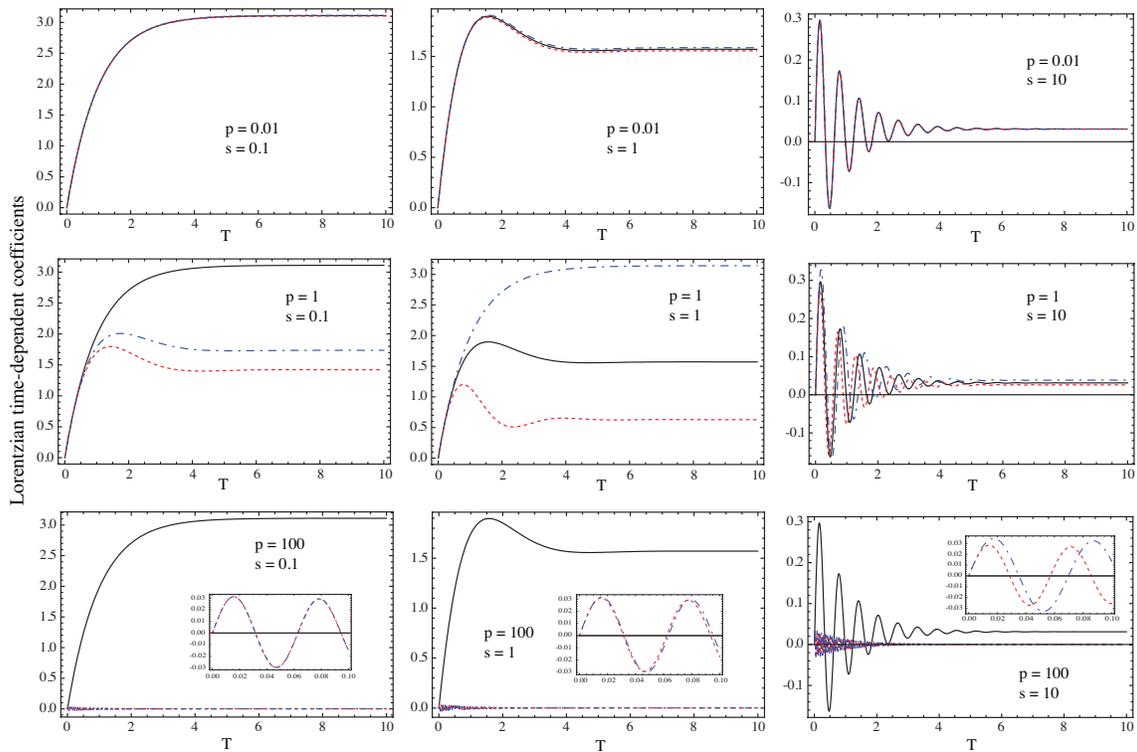}
\caption{(Colors online) Lorentzian time-dependent coefficients $\gamma_+(T)/\alpha^2$ (dot-dashed blue line), $\gamma_-(T)/\alpha^2$ (dashed red line) and $\gamma_0(T)/\alpha^2$ (solid black line) for  $p=0.01, 1, 100$ and $s=0.1,1,10$. The insets show very short time scale dynamics.
\label{Fig1}}
\end{figure*}

\subsection{Lorentzian reservoir}\label{sec:3a}
As a first example we consider a Lorentzian spectral density characterizing, e.g., one quantized mode of the electromagnetic field inside a cavity,
\begin{equation}
\label{Jl}
J_{Lor}(\omega)=\frac{\alpha^2}{2\pi}\frac{\lambda^2}{(\omega-\omega_0)^2+\lambda^2},
\end{equation}
where $\omega_0$ is the frequency of the mode supported by the cavity and $\lambda$ is the width of the distribution quantifying leakage of photons through the cavity mirrors. The  reservoir correlation time is given by $\tau_{C}=\lambda^{-1}$. The coupling constant $\alpha^2$ has frequency dimensions; the limit of weak coupling between the system and the environment, assumed in this paper, is valid when $\alpha^2$ is smaller than the smallest relevant frequency in the system.

\subsubsection{Time-dependent coefficients}
The time-dependent coefficients for the two-level atom in a Lorentzian reservoir can be calculated explicitly using Eqs. (\ref{rates})-(\ref{lambs}), and are given by
\begin{eqnarray}
\label{ratel}
\!\!\! \gamma_{\xi}(T)\!\!\!&=&\!\!\!\frac{\alpha^2 }{2(1+q_{\xi}^2)}\left(1-e^{-T}\cos q_{\xi}T+e^{-T} q_{\xi} \sin q_{\xi} T\right),\\
\!\!\! \lambda_{\xi}(T)\!\!\!&=&\!\!\!\frac{\alpha^2 }{1+q_{\xi}^2} \left(- q_{\xi} +e^{-T} q_{\xi} \cos q_{\xi} T+e^{-T}  \sin q_{\xi} T\right),\nonumber\\
\end{eqnarray}
where $T=\lambda t$ and $q_{\xi}=s-\xi p$ with $\xi=\{-,0,+\}$. We introduce two important parameters
\begin{eqnarray}
p&=&\frac{\tau_C}{\tau_S} =\frac{ \omega }{ \lambda},\\
s&=&\frac{\omega_0-\omega_L}{\lambda}.
\end{eqnarray}
The former of these two parameters identifies the region of validity of the secular approximation, more precisely $p\gg1$ corresponds to the secular regime and $p\ll1$ corresponds to the nonsecular regime. 
\begin{figure}
\includegraphics[width=7cm]{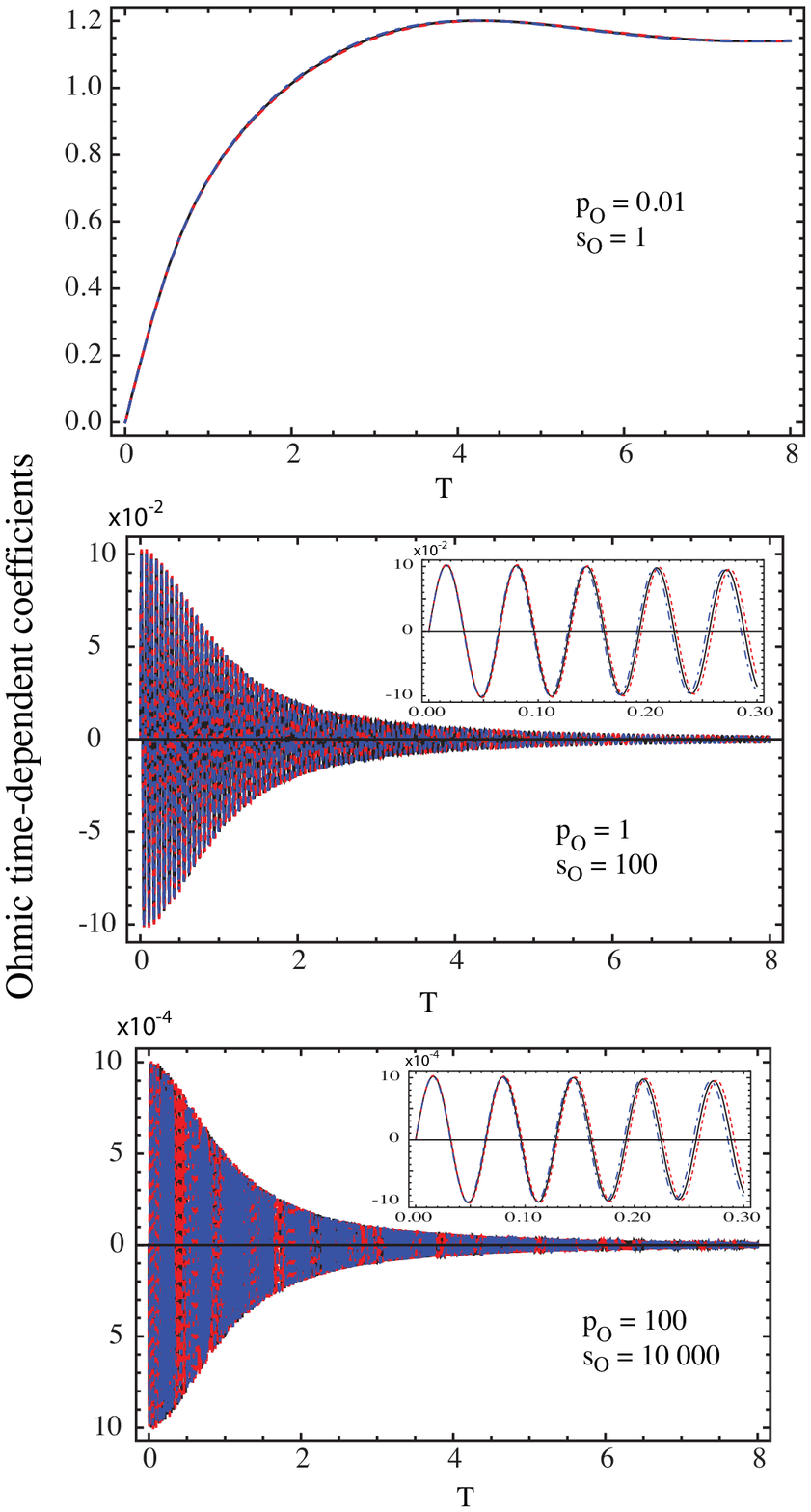}
\caption{(Colors online) Ohmic time-dependent coefficients $\gamma_+(T)/\alpha^2$ (dot-dashed blue line), $\gamma_-(T)/\alpha^2$ (dashed red line) and $\gamma_0 (T)/\alpha^2$ (solid black line) for $p=0.01, 1, 5$. The insets show very short time scale dynamics.}
\label{Fig2}
\end{figure}
The latter parameter $s=(\omega_0-\omega_L)/\lambda \approx (\omega_0-\omega_A)/\lambda$ indicates how far detuned is the peak of the Lorentzian spectrum from the atomic and/or laser frequency in units of $\lambda$.\\
 In Fig. 1 we plot, as an example, the dynamics of $\gamma_{\xi}(T)$  for different values of the parameters $p$ and $s$. We note in passing that the coefficient  $\gamma_0(t)$ does not depend on $p$ but only on $s$. A first look at the plots of Fig. 1 shows that for small values of $p$ (upper row), i.e., in the nonsecular regime, the time-dependent coefficients $\gamma_+(t)$, $\gamma_-(t)$ and $\gamma_0(t)$ coincide. Indeed, a power series expansion with respect to $p$ shows that $\gamma_{\pm}(t) \simeq \gamma_0(t)$ when $p\ll1$. In this case, however, the master equation contains the nonsecular dissipator of Eq. (\ref{dissipator'}) and hence it is not in time-dependent Lindblad form so we cannot describe the dynamics in terms of the NMQJ approach. We will numerically investigate the dynamics of this regime of parameters in Sec. IV. For intermediate and large values of $p$ the three coefficients are clearly distinct, as can be seen in the second and third row of Fig. 1. We now focus on the secular regime $p \gg 1$ described by the last row in Fig. 1.

\subsubsection{Non-Markovian Quantum Jumps}
In the secular regime the nonunitary dynamics is described only by the dissipator of Eq.  (\ref{secular dissipator}).  The decay rates associated to transitions between the dressed states are given by $C_{\pm}^2\gamma_{\pm}(t)$ while the decay rate associated to phase flips in the dressed state basis is given by $C_0^2\gamma_0(t)$. When the laser is resonant with the atomic transition, i.e., $\Delta=0$, then $C_+^2=C_-^2=C^2_0=1/4$. In this case Fig. 1 shows that, for all values of $s$, $\gamma_0(t)$ is the dominant decay rate so the main contribution to the system dynamics is given by phase flips in the dressed states.  A similar conclusion holds for $\Delta/\Omega \ll 1$, since in this case $C_+/C_0 \simeq C_-/C_0\simeq 1$.
On the other hand, for $\Delta/\Omega \gg 1$, one gets $C_+/C_0 \simeq 2 \Delta/\Omega$ and $C_-/C_0 \simeq 0$. In this case the dynamics is dominated by transitions between dressed states, in particular those described by the jump operator $\bar{\sigma}_+$, occurring at a rate $C_+^2 \gamma_+(t) \simeq  \gamma_+(t)$. \\
We note also that for increasing values of $s$ the stationary Markovian value of the time dependent coefficients decreases, due to a smaller effective coupling with the reservoir. Moreover, high values of $s$ are characterized by oscillatory behavior of all the three coefficients, independently from the value of $p$. For large enough values of $s$ the decay rates take temporarily negative values. This feature is typical of time-local non-Markovian open quantum systems and generally occurs when the characteristic frequency of the open quantum system, here $\omega_A \simeq \omega_L$, is detuned from  the peak of the reservoir spectrum \cite{Maniscalco06,Maniscalco04b}. Negative values of the decay rates are interpreted, in the NMQJ formalism, as reversed jumps canceling the jumps previously occurred in the same channel when the decay rate was positive. So the presence of oscillations around zero in the decay rates indicates non-Markovian dynamics induced by the reservoir memory and describing the feedback of information and/or energy from the reservoir into the system \cite{JPRL09}. It is worth noticing that non-Markovian oscillations occur in $\gamma_{\pm}(t)$ for all values of $s$, in the secular regime $p \gg 1$. \\
In the next subsection we present the analytic expressions of the same coefficients for a different reservoir spectral density, namely the Ohmic one, and study their time evolution. Comparing the behavior of $\gamma_{\xi}(t)$ for the two types of spectra we will see which features are common and which ones vary significantly when changing the spectrum.

\subsection{Ohmic reservoir}
We now focus on the Ohmic spectral density with exponential cut-off function
\begin{equation}
\label{Jo}
J_{Ohm}(\omega)=\alpha^2\,\omega\exp\left[-\frac{\omega}{\omega_C}\right],
\end{equation}
where $\omega_C$ is the cut-off frequency and $\alpha$ is a dimensionless coupling constant in the limit of weak coupling between the system and the environment, i.e., $\alpha\ll1$. The inverse of the cut-off frequency is the Ohmic reservoir correlation time $\tau_{C}=\omega_C^{-1}$.

\subsubsection{Time-dependent coefficients}
The non-Markovian time-dependent coefficients for the two-level atom in an Ohmic reservoir take the form
%\begin{widetext}
\begin{eqnarray}
\label{rateo}
\gamma_{\xi}(T)&=&\alpha^2\,\omega_C \Big\{ \frac{2}{1+T^2}\Big[T\cos\left(q_o T\right)+\sin\left(q_o T\right)\Big]  \\
&+& q_o e^{-q_o}\left(\pi-i\text{Ci}\bar{z}+i\text{Ci}z+\text{Si}\bar{z}+\text{Si}z\right) \Big\}, \nonumber\\
\lambda_{\xi}(T)&=&\frac{\alpha^2\omega_C}{2}\frac{1}{1+T^2}\Big\{ \Big[ \cos(q_o T)+T \sin(q_0 T)-1-T^2\Big]\nonumber\\
&+& q_o e^{-q_o}\Big[2\text{Chi}(q_o)+2\text{Shi}(q_o)-\text{Ci}\bar{z}-\text{Ci}z \nonumber \\
&+& i\text{Si}\bar{z}+i\text{Si}z\Big] \Big\}, \label{rateo2}
\end{eqnarray}
%\end{widetext}
where $T=\omega_C t$, $z=q_o (T+i)$, $q_o = s_O + \xi p_O$, and the parameters $s_O$ and $p_O$ correspond to the $s$ and $p$ parameters introduced in the previous subsection, but now adapted to the Ohmic reservoir spectral density, i.e., $s_O=\omega_L/\omega_C$ and $p_O=\omega/\omega_C$. 
Moreover, in Eqs. (\ref{rateo})-(\ref{rateo2}),  the bar is used to denote complex conjugation, $\text{Ci}$ and $\text{Si}$ are the cosine and the sine integrals and $\text{Chi}$ and $\text{Shi}$ are the hyperbolic cosine and hyperbolic sine integrals, respectively.\\
Similarly to the Lorentzian case of the previous subsection, $p_O\gg1$ corresponds to the secular regime and $p_O\ll1$ corresponds to the nonsecular regime. We note, however, that, in the Ohmic case, differently from the Lorentzian case, the parameters $p_O$ and $s_O$  are no longer independent. Our model of a driven two-level atom is valid when the Rabi frequency $\Omega$ and the detuning between the atom and the laser $|\Delta|=|\omega_A-\omega_L|$ are small compared to the atomic frequency $\omega_A$. This imposes a restriction on the relative values of $p_O$ and $s_O$, in particular we must have
\begin{equation}
\label{ }
p_O \ll s_O.
\end{equation}
The Ohmic time-dependent coefficients for the secular, nonsecular and intermediate regimes are shown in Fig. 2. As in the Lorentzian case, in the nonsecular regime, the three decay rates coincide. In this case, therefore, similar considerations as those done in Sec. \ref{sec:3a} apply.
Again, in the nonsecular and intermediate regimes, the master equation is not in the Lindblad form, hence little can be said about the dynamics from the behavior of the decay rates only. 

\subsubsection{Non-Markovian Quantum Jumps}
In the secular regime, $p\gg 1$, the Ohmic coefficients display oscillatory behavior, similarly to the Lorentzian  case of Fig. 1 (last row). For the Ohmic reservoir, however, all the three time-dependent coefficients are of similar order of magnitude. For short times they coincide, as one can see from the inset in Fig. 2, but as time passes they start oscillating out of phase. As a consequence, when the laser is resonant with the atomic transition, i.e., $\Delta=0$, both quantum jumps between dressed states and phase flip jumps contribute to the dynamics, contrarily to the Lorentzian case where the phase flips between dressed states were dominant.\\ 
Summarizing, in this section we have explored the differences in the dynamics due to different reservoir spectra. We have seen that for both the Lorentzian and the Ohmic spectral density, in the nonsecular regime, the three coefficients $\gamma_{\xi}(t)$ appearing in the Lindblad-type master equation have the same time dependency. \\
Nonetheless, different spectral distributions corresponding to different physical environments do give rise to noticeable differences. As an example, we have seen that, in the secular regime and in the case of resonance $\Delta=0$, the type of quantum jumps occurring in the systems do depend on the spectral properties: in the Lorentzian case phase flips between different eigenstates dominate the dynamics while in the Ohmic case  quantum jumps between different dressed states also occur. Moreover, not all the spectra are similarly compatible with the assumptions on which our model rely. The Ohmic reservoir, for example, imposes some limitations on the value of the physical parameters characterizing the dynamics.

\section{Bloch vector dynamics}\label{dynamics}
An alternative way to describe the dynamics of a driven two-state system is by means of the Bloch vector $\mathbf{R}(t)$ whose components are defined as
\begin{eqnarray}
R_i(t)&=&Tr[\rho(t)\sigma_i], \label{bloch vector}
\end{eqnarray}
with $i=x,y,z$.
The equation describing the dynamics of the Bloch vector, known as optical Bloch equation, can be obtained straightforwardly from Eq. (\ref{master equation})
\begin{equation}
\label{OBE}
\frac{d\mathbf{R}(t)}{dt}=[D(t)+D'(t)]\mathbf{R}(t)+\mathbf{d}(t)+\mathbf{d'}(t),
\end{equation}
with $D(t)+D'(t)$ the damping matrix and $\mathbf{d}(t)+\mathbf{d'}(t)$ the drift vector, whose explicit forms are given in Appendix A. Note that, also in this case, we separate the contribution of the nonsecular terms, contained in the primed quantities, from the contribution of the secular terms. As we will see in Sec. V, the optical Bloch equations prove to be particularly useful for studying the conditions under which the dynamical map is completely positive. Moreover, they provide us with a clear physical picture of the dynamics in terms of dephasing and dissipation phenomena, as described later in this section.

\subsection{Secular regime}
\begin{figure}
\includegraphics[width=7cm]{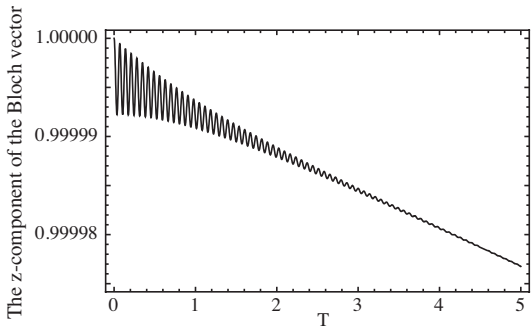}
\caption{Dynamics of the $z$-component of the Bloch vector as a function of $T=\lambda t$ for $p=100$ and $s=0.1$. We have set $\alpha^2 / \omega_A=0.01$ and $\Omega / \omega_A=0.01$.}
\label{Fig4}
\end{figure}
When $p\gg1$ the secular approximation is valid and  the dynamics of the $z$-component of the Bloch vector $R_z$, in the dressed state basis,  decouples from the $x$- and $y$-components. In this case the non-Markovian optical Bloch equations have a simple solution for any initial state $\mathbf{R}(0)=(x_0,y_0,z_0)$:
\begin{eqnarray}
\!\!\!\!R_x(t)\!\!\!&=&\!\!\!\exp[-\Gamma(t)](x_0\cos\omega t-y_0\sin\omega t),\label{rxsec} \\
\!\!\!\!R_y(t)\!\!\!&=&\!\!\!\exp[-\Gamma(t)](y_0\cos\omega t+x_0\sin\omega t),\label{rysec} \\
\!\!\!\!R_z(t)\!\!\!&=&\!\!\!e^{-\Lambda(t)}\!\! \left\{\!z_0\!+\!\!\!\int_0^t \!\!\!\! ds  e^{\Lambda(s)}\!\left[ C_-^2 \gamma_-(s)-C_+^2 \gamma_+(s) \right]\!\! \right\}  \label{highQsolution}
\end{eqnarray}
where
\begin{eqnarray}\label{Gamma}
\Gamma(t)\!\!\!&=&\!\!\!\frac{1}{2}\int_0^t ds \left[C_+^2\gamma_+(s)+C_-^2\gamma_{-}(s)+4C_0^2\gamma_0(s)\right],\\
\Lambda(t)\!\!\!&=&\!\!\!\int_0^t ds \left[ C_+^2\gamma_+(s)+C_-^2\gamma_{-}(s)\right], \label{Lambda}
\end{eqnarray}
and the time dependent coefficients $\gamma_{\xi}(t)$, with $\xi=\{+,0,- \}$, are given by Eq. (\ref{rates}). \\
It should be stressed that the solution of the non-Markovian Bloch equation, given by Eqs. (\ref{rxsec})-(\ref{highQsolution}), is valid for any form of the spectral density function $J(\omega)$ and therefore it can be used to describe the dynamics of a driven two-level system in any structured reservoir in the secular regime and in weak coupling. \\
As an example of the dynamics, we plot in Fig. 3 the time evolution of the $z$-component of the Bloch vector for the initial state $\mathbf{R}(0)=(0,0,1)$ in the Lorentzian case. For times of the order of the reservoir correlation time the Markovian exponential decay of $R_z$ is replaced by rapid non-Markovian oscillations, occurring at the same frequency of the oscillations of  $\gamma_+(t)$ and $\gamma_-(t)$. Physically these oscillations correspond to a rapid exchange of energy and information between the two-level atom and the environment due to the reservoir memory.

\subsubsection{Markovian limit}
For times longer than the reservoir correlation time  $\tau_C$ the time-dependent decay rates $\gamma_{\xi}(t)$ approach their Markovian stationary values $\gamma_{\xi}^M$ and the solution of the Bloch equations reduces to the well known Markovian one \cite{cohen}
\begin{eqnarray} 
R_x(t)&=&e^{- t/{\tau_D}}(x_0\cos\omega t-y_0\sin\omega t),\label{Markovrxsec}\\
R_y(t)&=&e^{-t/\tau_D}(y_0\cos\omega t+x_0\sin\omega t), \label{Markovrysec}\\
R_z(t)&=&e^{-t/\tau_R}(z_0-z_{\infty})+z_\infty, \label{markovianhighQsolution}
\end{eqnarray}
where 
\begin{equation}
\label{zinf}
z_{\infty}=\frac{C_-^2 \gamma_-^M - C_+^2 \gamma_+^M}{C_-^2 \gamma^M_- + C_+^2 \gamma_+^M}
\end{equation}
is the $z$-component of the stationary Bloch vector $\mathbf{R}_{\infty}\equiv (0,0,z_\infty)$ and $\gamma^M_-$, $\gamma^M_+$ are the Markovian stationary values of $\gamma_-(t)$ and $\gamma_+(t)$, respectively. 
In Eqs. (\ref{Markovrxsec})-(\ref{markovianhighQsolution}), the Markovian relaxation time $\tau_R$ and decoherence time $\tau_D$ are
\begin{eqnarray}
\tau_R^{-1} &=&\frac{1}{2} \left[C_+^2\gamma_+^M+C_-^2\gamma_{-}^M+4C_0^2\gamma_0^M\right]  \\
\tau_D^{-1} &=& C_+^2\gamma_+^M+C_-^2\gamma_{-}^M.
\end{eqnarray}
When the driven two-state system interacts with a structured environment, the relaxation and decoherence rates become time-dependent, the time dependency being determined by the form of the reservoir spectrum. \\
The equations above show that, in the Markovian limit, the well known relationship $2 \tau_R \ge \tau_D$ is satisfied. A close inspection to Eqs. (\ref{Gamma})-(\ref{Lambda}) shows that, even in the non-Markovian case, the time dependent decay rates always satisfy the relation $2 \Gamma(t) \ge \Lambda (t) $, if the time dependent coefficients $\gamma_{\xi}(t)$ are positive at all times. We have seen however, that the time dependent coefficients may oscillate taking temporarily negative values (See Figs. 1 and 2). In this case it is not a priori guaranteed that the inequality   $2 \Gamma(t) \ge \Lambda (t) $ holds. Stated another way, at certain time instants, the generalized time-dependent relaxation and decoherence times may violate the $2 \tau_R (t) \ge \tau_D (t)$ inequality. We will explore in detail this issue in Sec. \ref{cp}.\\
It should be noted that, in the secular regime and in the limit of weak coupling between the system and the environment, the non-Markovian effects are small and occur at a time-scale that is many orders of magnitudes smaller than the relaxation time of the two-level system. However, as we will see in the following subsection, in the nonsecular regime, strong oscillations may characterise the dynamics of the Bloch vector also at longer time scales.

\begin{figure*}
\includegraphics[width=15 cm]{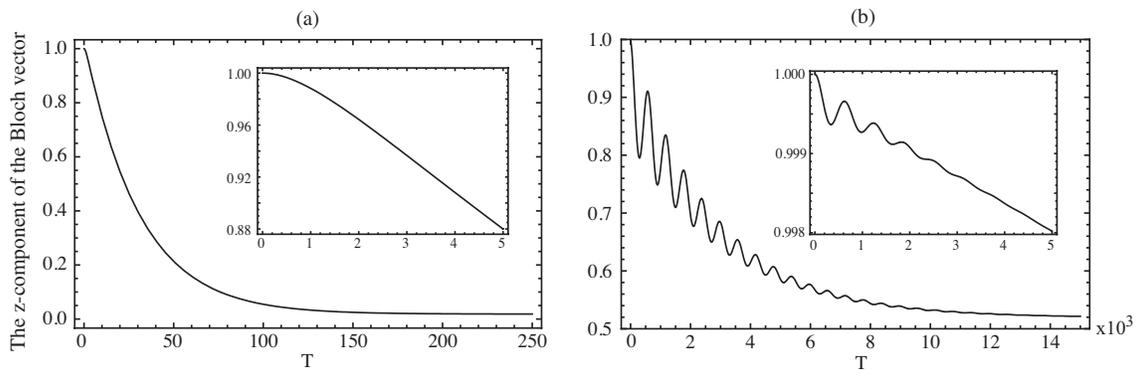}
\caption{The $z$-component of the Bloch vector in the Lorentzian case and in the nonsecular regime as a function of $T=\lambda t$ for $p=0.01$ and (a) $s=0.1$, (b) $s=10$. In each plot we have set $\alpha=0.01\omega_A$ and $\Omega=0.01\omega_A$. The insets show the dynamics in the short, non-Markovian time-scale.}
\label{Fig5}
\end{figure*}

\begin{figure*}
\includegraphics[width=15cm]{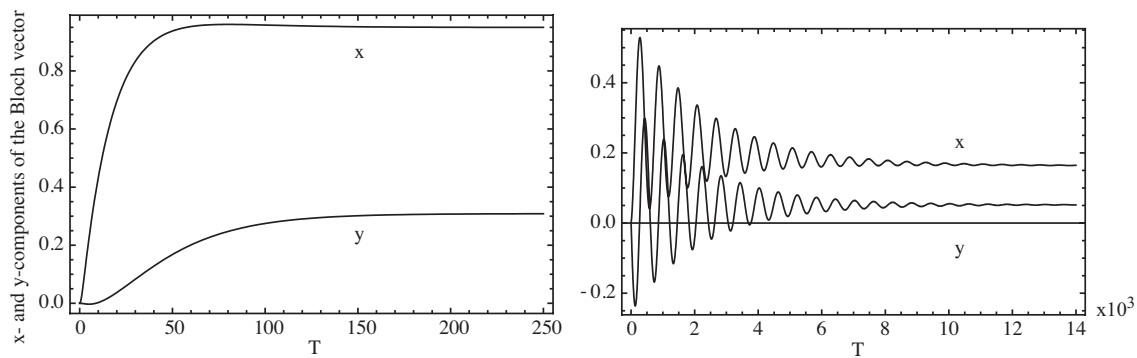}
\caption{The $x$ and $y$-components of the Bloch vector in the Lorentzian case and in the nonsecular regime as a function of $T=\lambda t$ for $p=0.01$ and (a) $s=0.1$, (b) $s=10$. In each plot we have set $\alpha=0.01\omega_A$ and $\Omega=0.01\omega_A$.}
\label{Fig7}
\end{figure*}

\subsection{Nonsecular regime}

When $p\ll1$ we cannot neglect the  nonsecular terms in the dynamics of the driven two-level system. Due to the presence of these terms the equation of motion for the $z$-component of the Bloch vector does not decouple anymore from the equations for the $x$- and $y$- components.
Therefore, we can no longer obtain an analytical solution for the optical Bloch equations. Furthermore, the master equation of the driven two-state system is no longer in the time-dependent Lindblad form and the microscopic description of the dynamics of the system in terms of the NMQJ method is not straightforward. \\ 
A numerical study of the solution of the full non-Markovian optical Bloch equations shows, however, substantial differences in the dynamics compared to the secular regime. 
In Fig. 4 we plot the time evolution of $R_z(T)$ for $p=0.01$ when the reservoir spectrum is Lorentzian, for two exemplary values of $s$.  Interestingly, while for small $s$ $R_z(T)$ decays monotonically, for large $s$ strong oscillations are present and last for long times. These oscillations are due to the nonsecular terms and have to be distinguished from the short-time non-Markovian oscillations. The latter ones, occurring at the correlation time $\tau_C$, are shown in the inset of Fig. 4 (b). \\
The behavior of $R_z(T)$ can be traced back to the dynamics of the time-dependent coefficients $\gamma_{\xi}(T)$. We recall that, in the nonsecular regime, these three coefficients coincide, i.e., $\gamma_{\pm}(T)=\gamma_0(T)\equiv \gamma(T)$.
As shown in Fig. 1 (first row), for $p=0.01$ and $s=0.1$, all three decay rates are positive hence we do not expect short time non-Markovian oscillations in the dynamics of $R_z(T)$. The initial quadratic decay of Fig. 4(a) is due to the fact that $\gamma(T)$, for $T \le 1$, is smaller then its Markovian stationary value and consequently the decay of $R_z(T)$ is slower than the one predicted by the Markovian theory. \\
For $p=0.01$ and $s=10$, on the contrary, Fig. 1 shows that $\gamma(T)$ oscillates taking negative values. These oscillations are responsible for the non monotonic behavior of  $R_z(T)$ at short times, see inset of Fig. 4 (b). The nonsecular oscillations occur on a time-scale comparable to the relaxation time. Similar oscillations can be seen in the dynamics of the $x$- and $y$-components of the Bloch vector, see Fig. 5(b). \\
A second difference with the secular dynamics, well visible from Fig. 5, is that the stationary states of  $R_x(t)$ and $R_y(t)$ are now no longer zero, at contrast with the prediction of the secular equations (\ref{rxsec})-(\ref{rysec}). In the bare state basis this corresponds to a non-zero steady state value of all three components of the Bloch vector, as one can see from Eq. (\ref{eq:RB}). The nonzero stationary value of  $R_z^B$, an effect known as vacuum-field dressed state pumping \cite{lewenstein}, has been observed experimentally in Ref. \cite{Zhu}. On the other hand, the nonzero stationary value of $R_x^B$ and $R_y^B$ indicates a stationary value of the atomic dipole moment different from zero, leading to substantial changes in the resonance fluorescence spectrum  \cite{lewenstein}.

\section{Complete positivity}\label{cp}
Theoretical descriptions of non-Markovian open quantum systems are often based on a series of assumptions and approximations without which it would not be possible to tackle the problem of the description of the dynamics in simple analytic terms. In the case described in the paper, e.g., the main assumptions and approximations are the factorized initial condition, i. e., $\rho(t=0)=\rho_S(t=0) \otimes \rho_R(t=0)$, with $\rho_S$ and $\rho_R$ the system and reservoir density matrices, respectively, the weak coupling approximation and, in some cases, the secular approximation. \\
The non-Markovian master equation we have used in the paper is not in Lindblad form, since even in the secular regime, the time-dependent coefficients $\gamma_{\xi}$ may temporarily take negative values. Therefore, both positivity and complete positivity (CP) of the dynamical map that, for Markovian systems in Lindblad form, are automatically guaranteed by the Lindblad-Gorini-Kossakowski-Sudarshan theorem \cite{lindblad, gorini}, can here be violated indicating that our solution no longer describe a physical state of the system. \\
In this section we will present the first study of complete positivity for the non-Markovian driven two-state model. We will derive explicit conditions for CP, and therefore positivity, of the dynamical map and we will see how these conditions have a clear and important physical interpretation. Once again, in the following subsections we will distinguish between the secular and nonsecular regimes.

\subsection{Secular regime}
Let us begin with the case in which the secular approximation is valid and we can neglect the nonsecular damping matrix $D'(t)$ and the nonsecular drift vector $\mathbf{d}'(t)$ in Eq. (\ref{OBE}). In this case the  damping matrix is in  block diagonal form, see Eq. (\ref{damp}). In the secular regime we can directly use  the CP conditions presented in Ref. \cite{hall}. The details of the calculation are presented in Appendix B.\\
We find that the necessary and sufficient condition for CP for the driven two-state system, in the secular regime and in weak coupling, is given by
\begin{equation}
\label{NScond}
2\Gamma(t)\geq \Lambda(t)\geq0.
\end{equation}
Note that the physical meaning of Eq. (\ref{NScond}) is straightforward since the inverse of $\Gamma(t)$ and $\Lambda (t)$ are the non-Markovian deoherence and relaxation times respectively. Therefore the necessary and sufficient condition for CP, in the secular approximation, is that the decoherence rate is at each time instant twice as big as the relaxation rate. In the Markovian limit the conditions of Eq. (\ref{NScond}) reduces to the well known condition
\begin{equation}
\label{markoviannecc}
2\,\tau_R\geq\tau_D\geq0.
\end{equation}
Recall from Section IV A that the Markovian condition is always satisfied by the driven two-state system for any spectral density.
Interestingly, inserting Eqs. (\ref{Gamma})-(\ref{Lambda}) into Eq. (\ref{NScond}) one sees that the necessary and sufficient condition for CP is equivalent to
\begin{equation}
\int_0^t ds\, \gamma_0(s) \ge 0. \label{SecCP}
\end{equation} 
It is worth noting that the condition above does not depend on $\omega=\sqrt{\Delta^2+\Omega^2}$, as one can see from Eq. (\ref{rates}).

\subsection{Nonsecular regime}
The method used to study the CP condition in the secular regime is no longer applicable in the nonsecular regime, when the damping matrix is not in a block diagonal form anymore. We use a more general method, based on the positivity of the Choi matrix, and make use of the weak coupling limit \cite{cresser}. Again the details of the calculation are given in Appendix B.\\
We find that the necessary and sufficient condition for CP for the driven two-state system, in the nonsecular regime and in weak coupling, is given by
\begin{equation}
\Lambda(t)+2\Gamma(t)\geq0.
\end{equation}
We note in passing that in the nonsecular regime $\Lambda(t)$ and $\Gamma(t)$ cannot be interpreted anymore as decoherence and relaxation non-Markovian rates, since the form of the master equation is now much more complicated and no simple analytical solution for the optical Bloch equation can be found. However, recall that in the nonsecular regime the three decay rates coincide, i.e., $\gamma_\pm(t)=\gamma_0(t)\equiv\gamma(t)$. Therefore, we find that the CP condition takes a form very similar to the one valid for the secular regime, given by Eq. ({\ref{SecCP}}),
\begin{equation}
\label{ }
\int_0^tds\,\gamma(s)\geq0.
\end{equation}

\section{Conclusions}\label{conclusions}
In this paper we have studied the non-Markovian dynamics of a driven two-state system immersed in a structured environment. We have derived the non-Markovian master equation and the optical Bloch equations both with and without the secular approximation, and we have presented the solution in terms of the Bloch vector dynamics for general reservoir spectra. \\
We have compared the dissipative dynamics of the driven two-state system for two different reservoirs, namely the Lorentzian and the Ohmic reservoir, and we have discovered that it is strongly influenced by the spectral properties. For example, in the secular regime and on resonance, in the Lorentzian case the dynamics is dominated by phase jumps in the eigenstate basis, while in the Ohmic case the dominant quantum jumps describe transitions between the dressed states. \\
We have discovered the existence of strong and long living nonsecular oscillations in all components of the Bloch vector in some regions in the parameter space. The nonsecular terms were also discovered to have a significant effect on the stationary quantum state of our system. An interesting open question we will consider next is whether the nonsecular oscillations describe a feedback of information/energy from the system into the environment as measured, e.g., by the non-Markovianity measure proposed in Ref. \cite{JPRL09}.  \\
We have also studied the validity of the secular approximation and how it depends on the spectral properties. In particular, our results show that in the Ohmic reservoir the use of the secular approximation is more subtle than in the Lorentzian case. That is, in the Ohmic case, one cannot always perform the secular approximation whenever $p_O\ll1$, since the model is not valid for $p_O\gtrsim s_O$, but instead both conditions have to be met before the secular approximation can be applied. \\
Finally, we have investigated in detail the complete positivity condition in both the secular and the nonsecular regime. We have discovered that this condition can be traced back to the behavior of the time-dependent coefficients appearing in the master equation and proportional, in the secular regime, to the decay rates of the system. Moreover, we have discovered that whenever the system is in time-dependent Lindblad form, i.e., when the secular approximation is valid, the CP condition consists of an inequality linking the non-Markovian decoherence and relaxation rates. Such inequality is the non-Markovian generalization of the well known condition $2 \tau_R \ge \tau_D$.\\
The dissipative driven two-state system is one of the most fundamental models of the theory of open quantum systems. Since most of our results are independent from the specific form of the spectral distribution, they can be straightforwardly applied to many different physical contexts where non-Markovian approaches are necessary, e.g., for implementations of quantum computing and other quantum technologies.\\

\begin{acknowledgments}
This work was supported by the Emil Aaltonen foundation, the Finnish Cultural foundation, and by the Turku Collegium of Science and Medicine (S.M.). We acknowledge stimulating discussions with J. Piilo.\\
\end{acknowledgments}
\appendix

\section{Non-Markovian optical Bloch equations}
In section IV we introduced the optical Bloch equations describing the dynamics of the driven two-state system. More explicitly the damping matrix and the drift vector in the optical Bloch equations of Eq. (\ref{OBE}), expressed in terms of the time-dependent decay rates, are as follows:
\begin{widetext}
\begin{equation}
\label{damp}
D(t)=\left(\begin{array}{ccc}-\frac{1}{2}\left[C_+^2\gamma_+(t)+C_-^2\gamma_{-}(t)+4C_0^2\gamma_0(t)\right]& -\omega & 0 \\ \omega &-\frac{1}{2}\left[C_+^2\gamma_+(t)+C_-^2\gamma_{}(t)+4C_0^2\gamma_0(t)\right] & 0 \\0 & 0 & -C_-^2\gamma_-(t)-C_+^2\gamma_+(t)\end{array}\right),
\end{equation}
\begin{equation}
\label{damp'}
D'(t)=\left(\begin{array}{ccc}\frac{1}{2}C_+C_-\left[\gamma_+(t)+\gamma_-(t)\right] & C_+C_-\left[\lambda_+(t)-\lambda_-(t)\right] & C_0\left[C_-\gamma_-(t)+C_+\gamma_+(t)\right] \\C_+C_-\left[\lambda_+(t)-\lambda_-(t)\right]  & -\frac{1}{2}C_+C_-\left[\gamma_+(t)+\gamma_-(t)\right] & 2C_0\left[C_+\lambda_+(t)-C_-\lambda_-(t)\right] \\C_0\left[C_++C_-\right]\gamma_0(t) & 2C_0(C_--C_+)\lambda_0 & 0\end{array}\right),
\end{equation}
\begin{equation}
\label{drift}
\mathbf{d}(t)=\left(0,0,C_-^2\gamma_-(t)-C_+^2\gamma_+(t)\right),
\end{equation}
\begin{equation}
\label{drift'}
\mathbf{d}'(t)=\Big(C_0\left\{C_+[\gamma_0(t)+\gamma_+(t)]-C_-[\gamma_0(t)+\gamma_-(t)]\right\},2C_0\left\{C_-[\lambda_-(t)-\lambda_0(t)]+C_+[\lambda_+(t)-\lambda_0(t)]\right\},0\Big).
\end{equation}
\end{widetext}
Note that the optical Bloch equations are given in the dressed state basis of the driven two-level atom. The transformation between the Bloch vector in the dressed state basis $\mathbf{R}(t)$ and the Bloch vector in the bare state basis $\mathbf{R}^B(t)$ is given by
\begin{equation}
\label{eq:RB}
\mathbf{R}^B(t)=\left(\begin{array}{ccc}\cos\theta & 0 & -\sin\theta \\0 & 1 & 0 \\ \sin\theta & 0 & \cos\theta\end{array}\right)\mathbf{R}(t),
\end{equation}
where $\theta=\arctan(\Omega/\Delta)$, i.e., the change of basis amounts to a rotation of the Bloch vector.

\section{Complete positivity}
In Section V we give the necessary and sufficient conditions for the complete positivity of the dynamics of the driven two-state system. Here we describe in more detail the derivation of the CP conditions.
\subsection{Secular regime}
The necessary condition for CP, for the driven two-state system, is given by the following two inequalities \cite{hall}
\begin{eqnarray}
\Lambda(t)&\geq&0, \label{necc1}\\
2\Gamma(t)&\geq&\Lambda (t). \label{necc2}
\end{eqnarray}
The sufficient condition for CP is also given by two inequalities. The first one coincides with Eq. (\ref{necc2}) and the second one can be expressed in the form
\begin{equation}
\label{suff2}
1+\varphi(t)^2-\chi(t)-2|\varphi(t)-\chi(t)|-\psi(t)^2\geq0, 
\end{equation}
where we have introduced the following auxiliary functions
\begin{eqnarray}
\varphi(t)&=&e^{-\Lambda(t)}\label{aux1}\\
\chi(t)&=&e^{-2\Gamma(t)}\\
\psi(t)&=&R_z(t)-e^{-\Gamma(t)}z_0 \label{aux3},
\end{eqnarray}
When the condition (\ref{necc2}) holds then the second sufficient condition simplifies to
\begin{equation}
\label{sufsimp}
[1-\varphi(t)]^2+\chi(t)-\psi(t)^2\geq0.
\end{equation}
In the Markovian limit, having in mind 
Eq. (\ref{zinf}), one sees that Eq. (\ref{sufsimp}) is equivalent to requiring that the stationary Bloch vector is contained inside the Bloch sphere.\\
In the non-Markovian time-scale we make use of the fact that the master equation (\ref{master equation}), and the corresponding optical Bloch equation, are valid to second order in  the coupling constant $\alpha$. Expanding Eqs. (\ref{aux1})-(\ref{aux3}) with respect to $\alpha$  Eq. (\ref{suff2}) becomes
\begin{equation}
\label{nonmarkoviansuff}
1-2\Gamma(t)+\mathcal{O}(\alpha^4)\geq0.
\end{equation}
For large values of $t$, $\Gamma(t)$, given by Eq. (\ref{Gamma}), grows without bound. In the short non-Markovian time-scales we are interested in, however, Eq. (\ref{nonmarkoviansuff}) is always valid. \\

\subsection{Nonsecular regime}
In the nonsecular regime we use a method based on the positivity of the Choi matrix \cite{cresser}. For a two-level system whose dynamics is given by a master equations with dissipator $\mathcal{D}$, the Choi matrix is computed as follows:\\
(1) Compute the auxiliary matrix $L$ defined by
\begin{equation}
\label{ }
L_{ij}=\frac{1}{2}Tr[\sigma_i\mathcal{D}(\sigma_j)],
\end{equation}
where $i,l\in\{0,1,2,3\}$, $\sigma_0$ is the identity matrix and we number the Pauli matrices as $\sigma_{1,2,3}=\sigma_{x,y,z}$, respectively.\\
(2) Define a second auxiliary matrix $F$ as
\begin{equation}
\label{ }
F(t)=\mathcal{T}\exp\left[\int_0^tds\,L(s)\right],
\end{equation}
where $\mathcal{T}$ is the time-ordering operator.\\
(3) The Choi matrix is now defined as
\begin{equation}
\label{ }
S_{ab}=\frac{1}{4}\sum_{i,j=0}^{3}F_{ij}Tr[\sigma_j\sigma_a\sigma_i\sigma_b],
\end{equation}
with $F_{ij}$ matrix elements of $F(t)$.
The dynamics of the two-level system is completely positive if and only if the Choi matrix $S$ is positive semi-definite, i.e., its eigenvalues are positive.\\
Using the fact that our master equation is valid to second order in perturbation theory with respect to $\alpha$, we write the matrix $\tilde{L}(t)\equiv\int_0^tdsL(s)$ as
\begin{equation}
\label{ }
\tilde{L}=\tilde{L}_0+\alpha^2\tilde{L}_2,
\end{equation}
where $\tilde{L}_0$ is a matrix containing all elements of $\tilde{L}$ independent of $\alpha$ and $\alpha^2 \tilde{L}_2$ contains all elements of $\tilde{L}$ proportional to $\alpha^2$. Then
\begin{equation}
\label{ }
F(t)=\exp\left[\tilde{L}(t)\right]=\exp\left[\tilde{L}_0(t)\right]\left[I+\alpha^2\tilde{L}_2(t)\right]+\mathcal{O}(\alpha^3).
\end{equation}
The eigenvalues of the Choi matrix, calculated neglecting all the terms of order greater than the second  in $\alpha$, are
\begin{eqnarray}
\epsilon_{1,2}&=&0\nonumber\\
\epsilon_{3,4}&=&1\pm\sqrt{1-[\Lambda(t)+2\Gamma(t)]}.
\end{eqnarray}
The eigenvalues $\epsilon_{3,4}$ are real whenever $\Lambda(t)+2\Gamma(t)\leq1$. This condition is always satisfied in the short non-Markovian time-scale when $\alpha\ll1$. This ensures that $\epsilon_3 \ge 0$. The condition of non-negativity of the last eigenvalue, i.e., $\epsilon_4\geq0$, is satisfied whenever $\Lambda(t)+2\Gamma(t)\geq0$.

\end{document}